\documentclass[preprint, showpacs,preprintnumbers,amsmath,amssymb,nofootinbib]{revtex4}
\usepackage{amssymb}
\usepackage{natbib}
\usepackage{graphicx}
\usepackage{dcolumn}
\usepackage{bm}

\newcommand{\bea}{\begin{eqnarray}}
\newcommand{\eea}{\end{eqnarray}}
\begin{document}
\title{ Exploring the potentiality of  standard sirens to probe cosmic opacity at high redshifts}
\author{ Xiangyun Fu$^1$\footnote{corresponding author:  xyfu@hnust.edu.cn}, Jianfei Yang$^1$,  Zhaoxia Chen$^1$, Lu Zhou$^1$, and Jun Chen$^2$}
\affiliation{$^1$Institute of  Physics, Hunan University of Science and Technology, Xiangtan, Hunan 411201, China\\
$^2$School of Science, Kaili University,
Kaili, Guizhou 556011, China
}

\begin{abstract}
 In this work, using the Gaussian process, we explore the potentiality of future gravitational wave (GW) measurements to probe cosmic opacity at high redshifts through comparing its opacity-free luminosity distance (LD) with the opacity-dependent one from the combination of Type Ia supernovae (SNIa) and gamma-ray bursts (GRBs). The GW data, SNIa and  GRB data are simulated from the measurements of the future Einstein Telescope,  the actual Pantheon compilation and the latest observation of GRBs compiled by L. Amati {\it et al}, respectively.  A nonparametric method is proposed to probe the spatial homogeneity of cosmic transparency  at high redshift by comparing the  LD reconstructed from the GW data with
 that reconstructed from the Pantheon and  GRB data. In addition,  the cosmic opacity is tested by using the parametrization for the optical depth, and the results show that the constraints on cosmic opacity are more stringent than the previous ones. It shows that the future GW measurements may be used as an important tool  to probe the cosmic opacity in the high redshift region.

$\mathbf{Keywords:}$  Cosmic opacity,  gravitation wave, Type Ia supernovae, gamma-ray bursts
\end{abstract}


 \maketitle

\section{Introduction}
In 1998, the evidence of the accelerating expansion of the Universe was first revealed by the unexpected dimming of the type Ia supernovae (SNIa)~\cite{Riess1998,Perlmutter1999}. In the frame of the General Relativity, a cosmic distribution of an exotic component with negative pressure, dubbed as dark  energy, has been suggested to explain the present acceleration.  On the other hand, a cosmological distribution of dust has been proposed to be an alternative explanation for this dimming phenomenon~\cite{Aguirre1999}. Indeed, the photons may be absorbed or scattered by dust in the Milky Way, intervening galaxies, the intergalactic medium, and their host galaxies~\cite{Lima2011}. However, it has always been a controversial topic that whether the extinction effect of SNIa has a great impact on the conclusion of cosmic accelerating expansion~\cite{Lima2011,Kang2020,Rose2020,Tutusaus2017,Huillier2019}. In addition, some other plausible mechanisms attempt to explain cosmic opacity, such as scalar fields coupled nonminimally with the electromagnetic (EM) Lagrangian~\cite{Hees2014,Holanda2017,Holanda20161,Aguirre1999} or oscillation of photons propagating in extragalactic magnetic fields into light axions~\cite{Csaki2002,Avgoustidis2010,Avgoustidis2009,Jaeckel2010}.   Exotic mechanisms for cosmic opacity are not fully understood yet. Any changes in the photon flux during propagation towards the Earth will affect
the luminosity distance (LD) measurements of light sources.  As the cosmic acceleration rate and the cosmological parameters determined by the LD measurements are highly dependent on the dimming effect, cosmic opacity still needs to be investigated accurately in the era of precision cosmology.

 The general tests on  cosmic opacity have been performed by using  the  cosmic distance-duality relation (CDDR), which connects the LD $D_{\rm L}$  with the
angular diameter distance (ADD) $D_{\rm A}$ at the same redshift $z$ through the  identity~\cite{eth1933}: ${D_{\rm L}}{(1+z)}^{-2}/{D_{\rm A}}=1$.
 Provided that light travels always along null geodesics in a
Riemannian geometry and the number of photons  is conserved, this reciprocal relation holds true in whatever cosmology~\cite{ellis1971,ellis2007}.
  If Riemannian geometry is used as the tool to describe the spacetime of the Universe and a photon traveling along null geodesic is more fundamental than the conservation of the photon number~\cite{uzan2004,Santana2017}, any violations of  the CDDR likely indicate the opacity of the Universe.
  Many works have been devoted to probe cosmic opacity with the SNIa, baryon acoustic oscillation (BAO),  the galaxy cluster samples, the Hubble data, the old passive galaxies, and the gas mass fraction of galaxy clusters~\cite{More20092, Avgoustidis2010,Avgoustidis2009,Liao2016,Jesus2016,Holanda20172,Liao2011,
  Liao2015,Li2013}.  No significant cosmic opacity is obtained in these studies.

It should be noted that most of the measurements are using EM radiation, although the cosmological parameters can be measured precisely from various astronomical observations.  Schuts first proposed that~\cite{Schutz},  using the fact that the waveform signals of GWs from inspiraling and merging compact binaries encode distance information, the Hubble constant can be determined from GW observations. So, this type of GW sources can be considered as standard sirens in astronomy, analogous to SNIa standard candles.   Compared with the distance estimation from  SNIa measurements,  the LDs can be obtained directly from the GW signals without the need of a cosmic distance ladder since stand sirens are self-calibrating. It is more important that GWs are able to propagate freely through a perfect fluid without any absorption and dissipation in the Friedman-Lema{\^i}tre-Robertson-Walker metric. Therefore, most objects (e.g. the Earth) are
nearly transparent to GWs~\cite{Weihao2019} and the standard siren can be considered as an opacity-free distance indicator. Generally speaking, it is hard to obtain the redshift of the GW source. But the source redshift might be obtained from the observation of EM counterparts that occur coincidentally with the GW events~\cite{Zhao2011, ET, Nissanke2010,Cai2017}, if compact binaries are neutron star (NS)-NS or black hole (BH)-NS binaries. Recently, the joint detection of GW event GW170817 with EM counterpart (GRB 170817A) from the merger of binary NSs~\cite{Abbott,Abbott2,Daz,Cowperthwaite} opened a new era of multi-messenger cosmology. Thus, the LD-redshift relation can be constructed by a cosmological model-independent way, and it can be used to make constraints on the parameters of cosmology.

The Einstein Telescope (ET), which is the third generation of the ground-based detector, will be able to detect GW signals ten times more sensitive in amplitude than the advanced ground-based detectors, and it can cover a wide range of frequency $1\sim 10^4$ Hz up to redshift $z\sim2$ for the NS-NS  and $z>2$ for the BH-NS mergers systems.  So, the measurements from ET will provide us with an opportunity to make constraint on the parameters of cosmology at relative higher redshifts.
Up to now, the simulated GW data have been used to measure the cosmological parameters~\cite{Pozzo201217,Liao2017,Cai2016,Wei2017,Belgacem2019,Jin2000,Agostino,ZhangJ2019,YangW2019}, determine the total mass of neutrino~\cite{Wang2018}, investigate the anisotropy of the Universe~\cite{Cai2018,Wei2018}, constrain the evolving Newton's constant G~\cite{Zhao2018}, discuss the estimation of  the Hubble parameter with the actual background expansion model of the universe~\cite{Shafieloo2020,Keeley2020}, and test the CDDR~\cite{Yang2017,fuxiangyun2019,Liaokai}.

More recently,  confronting the LD from Joint Light Analysis (JLA) SNIa or Pantheon compilation with the opacity-free LD from future observational GW data, Wei~\cite{JUN-JIE WEI2019}, Qi~{\it et al}.~\cite{Jing-Zhao Qi2019}, and Zhou~{\it et al}.~\cite{fuxiangyun20192} performed successively  unbiased tests on the cosmic opacity  with different methods. The results showed that future GW measurements would be at least competitive with current tests on cosmic opacity.
However, regarding the tests referred above, it is important to note that due to the limitation of the redshift distribution observed by SNIa, the tests are only limited to  the redshift range $0<z<2.0$. The cosmic opacity in the redshift region between the SNIa and cosmic microwave background (CMB) observations remains to be explored.  On the other hand, since current astronomical observation involving SNIa and other available data cannot yet confirm whether the density of dark energy is constant or time-varying, the LDs of gamma-ray bursts (GRBs) have been used to constrain cosmological parameters to determine the behaviors of dark energy at higher redshifts $2<z<8$. The measurements of GRBs can be considered as a new means to make constraint on the property of dark energy, once phenomenological relations are adopted to make the sources of bursts to be standardized candles~\cite{Amati,Schaefer,Samushia2010,Wei2013}. Thus, to have confidence in the analyses with the GRBs observations, it is important to test cosmic opacity in the high redshift region as well. Holanda {\it et al.} tried to probe the cosmic opacity with the LDs of SNIa, Hubble parameter $H(z)$ , and GRBs data at high redshifts ($z>1.4$) for a flat cosmological constant and
cold dark matter ($\Lambda$CDM) model~\cite{Holanda2014}, and found that the observational data were compatible with a transparent universe at $1\sigma$ confidence level (CL). However, it should be noted that the analyses in Ref.~\cite{Holanda2014} are model-dependent, thus some bias might result from some particular cosmological models. To draw firm and robust conclusion on the behaves of dark energy, one needs to employ new observational data and methods to probe the cosmic opacity by creasing the redshift depth. Therefore, future GW measurement and GRB observation  at high redshifts will be reliable candidates for this task, which is also the motivation of this paper.

In this work,  we explore the potentiality of future GW measurements to probe the cosmic opacity at high redshifts through comparing its opacity-free luminosity distance (LD)  with the opacity-dependent one from the combination of SNIa and GRBs data. GW data points are simulated from the ET,  the SNIa  data are simulated from the Pantheon compilation~\cite{Scolnic2018}, and GRB data are simulated from the latest catalogue which are obtained with a cosmological model-independent technique to overcome the circularity problem\cite{Amati2019}.  To match the GW data with the combination of SNIa and GRB data at the same redhsift, we  employ the Gaussian process~\cite{Seikel2012,Rasmussen2006} to reconstruct the continuous LD function from the mock GW measurements or the combination of SNIa and GRB data.
In our analysis, we first probe  the spatial
homogeneity of the cosmic transparency by using non-parametric method. Then, we test the cosmic opacity by adopting two types of parameterizations for the optical depth.   Compared to previous results,  our analyses show that future measurements of GW will be an important tool to probe cosmic opacity at high redshifts.

\section{data and the Gaussian process}
To dodge the impacts of any systematics from the combination of SNIa and GRB datasets, we compare the opacity-free LD from the mocked measurements of future GW events with the observed one obtained from the combination of `Pantheon-like' SNIa and `GRBs-like' compilations.

\subsection{Pantheon  SNIa  and GRB data}
 The  Pantheon SNIa compilation is released by  the Pan-STARRS1 Medium Deep Survey, which  consists of 1048   data points up to redshift $z\sim 2.26$. The LDs of the Pantheon compilation are calibrated from the SALT2 light-curve fitter through applying   the Bayesian Estimation Applied to Multiple Species with Bias Corrections method to determine
the nuisance parameters and taking account of the distance bias corrections. The distance modulus and its uncertainty  can be obtained from $\mu=m_{\rm B}-M_{\rm B}$ and $\sigma_{\mu}=\sigma_{m_{\rm B}}$, where $m_{\rm B}$ and $M_{\rm B}$ is the rest-frame peak magnitude in the B band and the absolute
magnitude of SNIa, respectively. Then the LD ($D_{\rm L}$)  can be derived from the relation
\begin{align}
\mu(z)=5\log_{10}[D_{\rm L}(z)]+25,
\label{equa:mu1}
\end{align}
and its uncertainty $\sigma_{D_{\rm L}}={\ln{10}}~\sigma_\mu D_{\rm L}$/5.

The GRBs can be observed up to redshift $z\sim 10$ due to the intense explosions in the Universe. An important observational aspect of long GRB is the several correlation between the spectral and intensity properties, which suggests that the GRB measurement can be used as a complementary cosmic probe to the standard candles~\cite{Schaefer,Bromm,Amati,Norris,Fenimore,Schaefer2003,
Ghirlanda,Liang2005,Firmani,Yu2009, Demianski2016}, although the mechanism behind GRBs explosions is not completely known yet. One of the most successful proposals is the Amati relation ($E_{\rm p}$-$E_{\rm iso}$ relation)~\cite{Amati}, which relates the rest frame spectral peak energy $E_{\rm p}$ and the bolometric isotropic-equivalent radiated energy $E_{\rm iso}$. However, the application of GRB observations for cosmology is also affected by the so-called circularity problem~\cite{Amati,Ghirlanda,Demianski2016}, which arises from the fact that, given the lack of low redshift measurements of GRBs, the $E_{\rm p}$-$E_{\rm iso}$ correlation is obtained through assuming a background cosmology. For example, using the standard $\Lambda$CDM model to calibrate the GRB measurements, it will inevitably return an overall agreement with the concordance model to make constraints on cosmological parameters of any dark energy framework.
Thus, calibrating the Amati relation is a challenge if one dodges the problem of circularity.
Recently,  Amati {\it et al.} proposed a model-independent technique to overcome the circularity problem with the Amati relation~\cite{Amati2019}, and built up a compilation consisting of 193 GRB data points. To calibrate this relation, they  fitted the most recent Hubble data with a B$\acute{\rm e}$zier parametric curve, and  employed the Hubble data to approximate the cosmic evolution. Then, with the combination of this GRB data set and SNIa JLA compilation, they made  observational constraints on the flat $\Lambda$CDM model, and  obtained that $(\Omega_{\rm m}, \alpha, \beta, {\rm M}_{\rm B})=(0.397\pm_{0.039}^{0.040},0.126\pm_{0.012}^{0.011}, 2.610\pm 0.130,19.090\pm 0.037)$ at $2\sigma$ CL.  Here,
 $\Omega_{\rm m}$  denotes the present
dark matter density parameter. The constraints on the cosmological parameters are obviously different from the results of current observations~\cite{Scolnic2018,SDSS2014}.

It should be noted that, due to the little knowledge of the physical processes driving the explosion, GRB observations, as the cosmological probing tools,  are not considered as important as the other astronomical measurements, such as SNIa, BAO, and CMB. There is a strong debate in Ref.~\cite{Heussaff2013} concerning whether the Amati relation is an intrinsic property of GRBs or merely a combination of selection effects. Moreover, Ref.~\cite{Salvaterra} studied the strong evolution in the LD function of GRB, and the results showed that GRBs might be intrinsically more luminous at high redshifts. If the Amati relation is used to calibrate the high-redshift GRB with the low-redshift ones, a smaller distance to GRB would be obtained.

Since the main purpose of this work is to probe the potentiality of future  GW measurements, it can be assumed that the problems of the GRB explosion will be solved in the next few decades, and the GRBs observation can be used as a complementary tool for SNIa measurements.  In this work, in order to doge any systematic impact while combining the SNIa data with the GRB data,  SNIa and GRB data are simulated from a flat $\Lambda$CDM. 137 deta points from the GRB compilation in the redshift range $1<z<5$ are added to Pantheon SNIa data to test cosmic opacity. The redshifts and uncertainties are taken from the actual Pantheon~\cite{Scolnic2018} and GRBs compilation~\cite{Amati2019}.  The fiducial LDs ($D_{\rm L}^{\rm fid}$)  of the mock data are obtained from a known flat $\Lambda$CDM with the most recent Planck results~\cite{Plank2015},
\begin{align}
h_0=0.678,~~~\Omega_{\rm m}=0.308,
\label{equa:A1}
\end{align}
where $H_0=100 h_0{\rm km s^{-1}Mpc^{-1}}$ denotes the present Hubble constant.
The measurements of LD ($D_{\rm L}$) can be derived from the random normal distribution through the equation $D_{\rm L}=\mathcal{N}(D_{\rm L}^{\rm fid}, \sigma_{D_{\rm L}})$. The mock SNIa and GRB data refer to `SNIa-like' and `GRB-like' data, respectively. The simulated method has been applied in discussing the estimation of  the Hubble parameter with different background expansion model~\cite{Keeley2020}. Our simulated results are presented in the left panel of Fig.~\ref{mock1}.

\subsection{Simulated GW data}

We simulate the GW data based on the ET which is the third-generation of ground-based GW detector. The ET will be able to detect GW signals to be ten times more sensitive in amplitude than the advanced ground-based detectors.  It can cover a wide range of frequency $1\sim 10^4$ Hz up to redshift $z
\sim5$.  The strategy implemented with the future GW detectors has been discussed in Refs.~\cite{Regimbau2012,Regimbau2014,Regimbau2015, Regimbau20152,Regimbau2016,Regimbau2017}.  In the following of this subsection, we will summarize  the process of mocked GW data for simplification.

Given that the effect, which is from the spin of the binary system and the change of orbital frequency over a single period, is negligible, for the waveform of GW, one can apply the stationary phase approximation to compute the Fourier transform $\mathcal{H}(f)$ of the time domain waveform $h(t)$,
\begin{align}
\mathcal{H}(f)=\mathcal{A}f^{-7/6}\exp[i(2\pi ft_0-\pi/4+2\psi(f/2)-\varphi_{(2.0)})],
\label{equa:hf}
\end{align}
where $\psi$ denotes the
polarization angle, $\varphi_{(2.0)}$ denotes the phase parameter, and the Fourier amplitude $\mathcal{A}$ is given by
\begin{align}
\mathcal{A}=&~~\frac{1}{D_{\rm L}}\sqrt{F_+^2(1+\cos^2(\iota))^2+4F_\times^2\cos^2(\iota)}\nonumber\\
            &~~\times \sqrt{5\pi/96}\pi^{-7/6}\mathcal{M}_{\rm c}^{5/6}\,.
\label{equa:A}
\end{align}
 Here, $\mathcal{M}_{\rm c}$ presents the chirp mass, $F_{+,\times}$ denotes  the beam pattern functions, ${D_{\rm L}}$ is the LD from GW signals,
 $t_0$ presents the epoch of the merger,  and $\iota$ responds to the angle of inclination.   For the simulation estimation of LD, we use the flat $\Lambda$CDM as the fiducial cosmological  model with the model parameters in Equ.~\ref{equa:A1}.

 The signal-to-noise ratio (SNR) for the network from the three independent ET interferometers can be written as
\begin{align}
\rho=\sqrt{\sum_{i=1}^{3}\langle\mathcal{H}^{(i)},\mathcal{H}^{(i)}\rangle}\,.
\label{equa:SNR}
\end{align}
Here, the inner product indicates the following equation
\begin{align}
\langle a,b\rangle=4\int_{f_{\rm lower}}^{f_{\rm upper}}{\widetilde{a}(f)\widetilde{b}^{\ast}(f)+\widetilde{a}^{\ast}(f)\widetilde{b}(f)\over{2}}{df\over{S_h(f)}}\,,
\label{equa:SNR1}
\end{align}
 here ${S_h(f)}$ represents the one-side noise power spectral density characterizing the performance of a GW detector. $f_{\rm lower}$ represents the  lower cutoff frequency which  is fixed at 1 Hz, and $f_{\rm upper}$ does the upper cutoff one decided by the last stable orbit (LSO), and $f_{\rm upper}=2f_{\rm LSO} $, where $f_{\rm LSO}=1/(6^{3/2}2\pi M_{\rm obs})$ is the orbit frequency at the last stable orbit. In this work,  the masses of NS and BH  are simulated with uniform distribution in the intervals $[1,2] M_\odot$ and  $[3,10] M_\odot$, respectfully, where $M_\odot$ denotes the solar mass. The ratio between BH-NS and NS-NS  binary systems is taken to be nearly 0.03~\cite{Cai2018,Abadie2010a}. As presented in the Advanced LIGO-Virgo network~\cite{Abadie2010a,Zhao2011}, the observational signal might be identified as a GW events, only if the interferometers have a network SNR of $\rho>8.0$.

Using the Fisher information matrix, one can obtain the instrumental uncertainty of the measurement of GW LD.   It is  assumed that the LD ($D_{\rm L}$) is uncorrelated with the errors on the remaining GW parameters (the inclination angle $\iota=0$), $\mathcal{H}\propto D_{\rm L}^{-1}$, and the corresponding instrumental uncertainty  can be written as, $\sigma_{D_{\rm L}}^{\rm inst}\simeq{2D_{\rm L}/ {\rho}}\,$. Furthermore, the LD is also affected by the effect of the weak lensing, and the lensing uncertainty can be assumed as $\sigma_{D_{\rm L}}^{\rm lens}=0.05z D_{\rm L}$~\cite{Cai2017,Zhao2011,Belgacem2019}. Thus, the total uncertainty of GW LD measurements is taken to be,
\begin{align}
\sigma_{D_{\rm L}}&=\sqrt{(\sigma_{D_{\rm L}}^{\rm inst})^2+(\sigma_{D_{\rm L}}^{\rm lens})^2}\nonumber \\
&= \sqrt{\bigg({2D_{\rm L}\over {\rho}}\bigg)^2+(0.05zD_{\rm L})^2}\,.
\label{equa:errorall}
\end{align}

It is expected that the source redshift can be obtained from the identified EM counterparts from the NS-NS and BH-NS binary systems. The redshift distribution of GW sources is
taken from the form of~\cite{Zhao2011,Sathyaprakash2010}
\begin{align}
P(z)\propto{4\pi d^2_{\rm C}(z)R(z)\over{H(z)(1+z)}}\,,
\label{equa:distri}
\end{align}
where $H(z)$ represents the Hubble parameter from the fiducial cosmological model, $d_{\rm C}$ is comoving distance, and $R(z)$ is the merger rate of binary systems with following expression~\cite{Cai2017,Zhao2011},
\begin{equation}
R(z)=\begin{cases}
1+2z, & (z\leq 1) \\
\frac{3}{4}(5-z), & (1<z<5) \\
0, & (z\geq 5).
\end{cases}
\label{equa:rz}
\end{equation}

\begin{figure}[htbp]
\includegraphics[width=8cm]{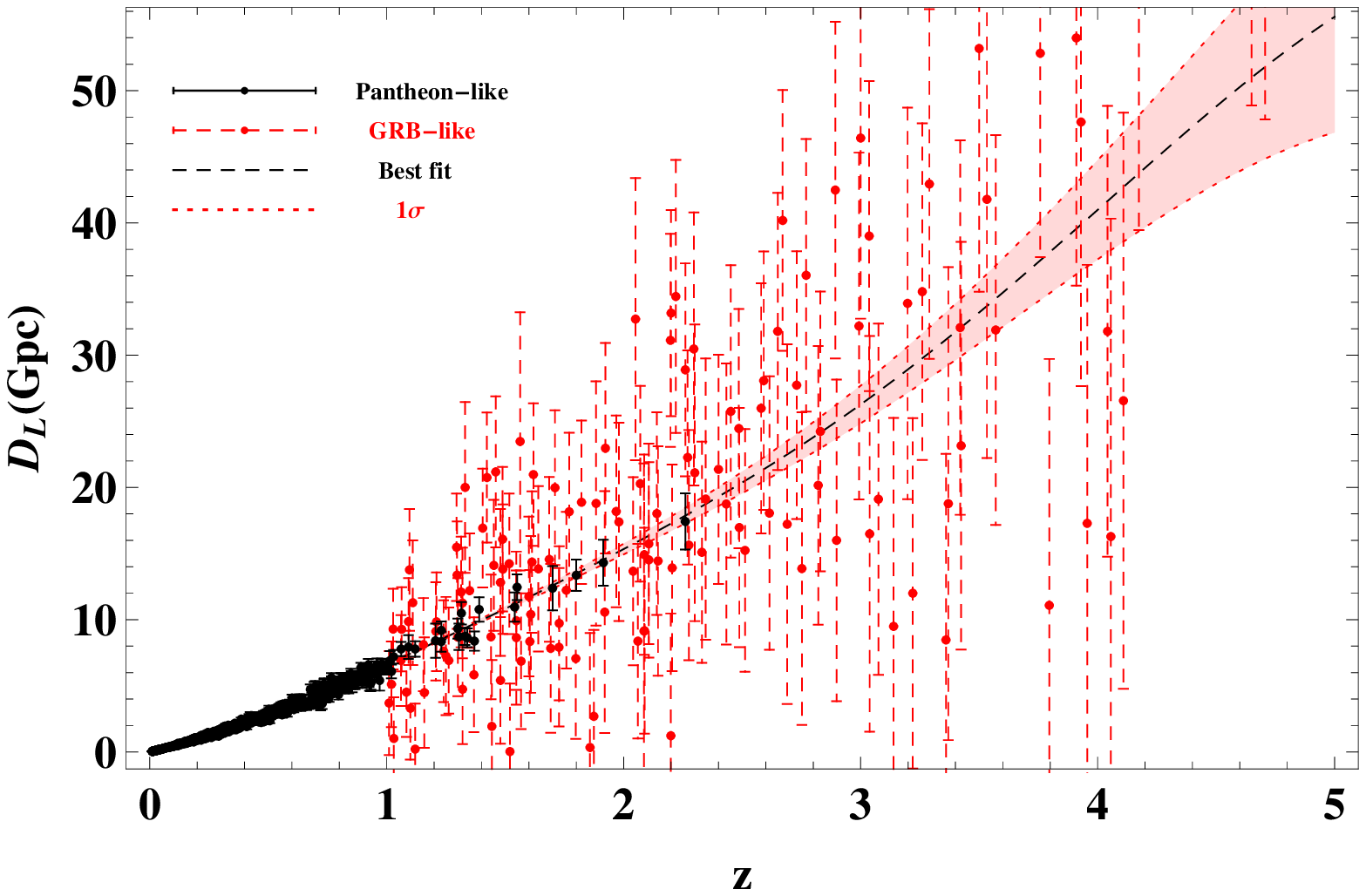}
\includegraphics[width=8cm]{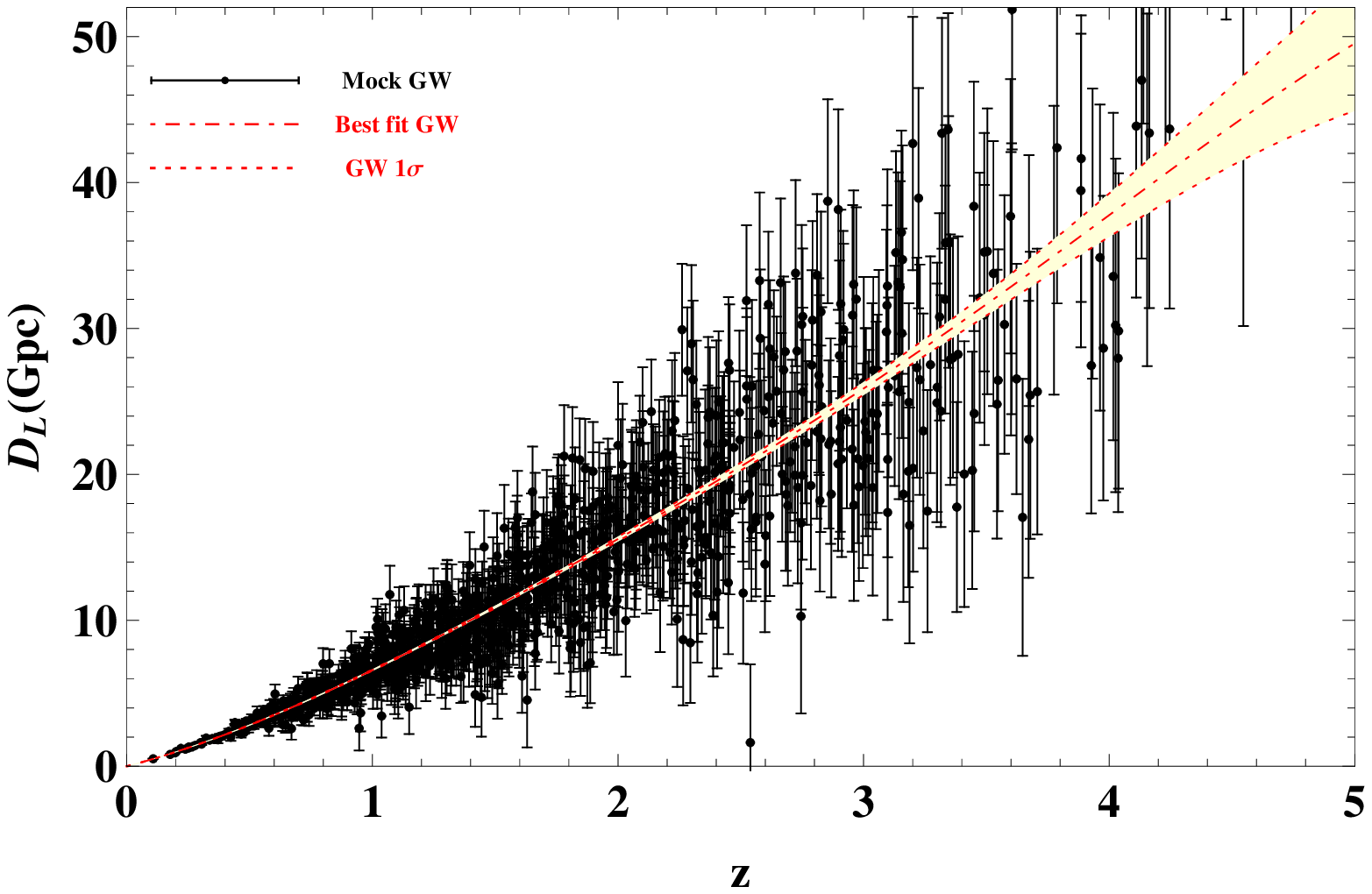}
\caption{\label{mock1} The sample catalogs of the `pantheon-like' data and `GRBs-like' data (left panel) , and  1000 mock GW events of redshifts (right panel). The corresponding smoothed function of LD obtained from  the Gaussian process method is also presented.}
\end{figure}

It is expected for the ET that the rates of  NS-NS and  BH-NS binary detections per year  are about the order $10^3-10^7$~\cite{ET}. However, this prediction about the rate is very uncertain. Then,  Cai and Yang~\cite{Cai2017,Yang2017} predicted that about $10^2$ GW measurements with EM counterparts will be observed per year, if the middle detection rate around $10^5$ is viable and the efficiency of EM counterparts are around $10^{-3}$ of the total number of binary coalescence. In this paper, following the process from Ref.~\cite{Zhao2011,Yang2017,Cai2017},
 we simulate 1000 data points in the redshift range $0<z<5$,
 and show the results in the right panel of Fig.~\ref{mock1}.

  \subsection{Gaussian process}

 In this work, we reconstruct the continuous function of LD from the mock GW or EM observational data with the Gaussian process. The advantages of the Gaussian process method are nonparametric and cosmological model-independent smoothing technique used to reconstruct a continuous function from the observed data.  The reconstructed function, at each redshift point $z$, is related to a Gaussian distribution with a mean value and Gaussian error bands.  The results from observational data points at any two redshifts $z_i$ and $z_j$ are correlated through a covariance function $\kappa(z_i,z_j)$ due to their nearness to each other. There is a wide range of possible candidates for this function, and it depends on  a set of hyperparameters. The different choices of the covariance function may affect the reconstruction to some
extent.  The detailed description and analysis of the Gaussian process can be found in Refs.~\cite{Seikel2012,Seikel20121}.  This method  has been widely used to reconstruct the equation of state of dark energy~\cite{Seikel2012,Cai2017}, and test cosmography~\cite{Shafieloo} and the CDDR~\cite{Yang2017,fuxiangyun2019}. The simplest covariance function usually takes the squared exponential as
\begin{equation}
\label{cov}
\kappa(z_i,z_j)=\sigma^2_f \exp\bigg[-{(z_i-z_j)^2\over{2 l^2}}\bigg],
\end{equation}
to implement the reconstructing process. Here, $\sigma_f$ and $l$ are two hyperparameters.  $\sigma_f$ denotes the output variance and fixes the overall amplitude of the correction in the $y$ direction, and $l$ gives the measure of the coherence length scale of the correlation in the $x$ direction. As discussed in Ref.~\cite{Keeley2020}, the hyperparameters play important role in determining the error bars of observation data. Both of the hyperparameters are optimized by the GP with the observed data set. In this work, we use the simplest squared exponential covariance function.  In order to avoid bias caused by different initial values, we change the initial value of the super parameter. After being optimized by the program, the returned values are about $\sigma_f\sim 34.48$ and $l\sim 2.49$.  The continuous functions of the LD from the mock  data  are shown in Fig.~1. It can be seen from this figure that, in the redshift region $2<z<5$, due to the small amount of GRB observation data (only $71$ data points) and its large error bar, the uncertainty of LDs reconstructed from the GRB data  is greater than that of LDs reconstructed from the GW measurements.

\section{Method}\label{method}
The LDs from future detectable GW sources are expected to be opacity-free. The LDs from the electromagnetic observations, i.e.  SNIa and GRB, are assumed to be systematically influenced by the cosmic opacity, as pointed out by Ref.~\cite{Avgoustidis2009}. The straightforward method to probe the cosmic opacity is to compare the observed LDs with the corresponding ones of GW measurements at the same redshifts.  The observed LD can be expressed as~\cite{More2009},
\begin{equation}\label{ddob}
  {D_{\rm L,obs}}(z)={D_{\rm L,true}}(z)e^{\tau(z)/2},
\end{equation}
if the photon flux received by the observers will be reduced by the factor $e^{-\tau(z)}$. Here $\tau(z)$ represents the optical depth related to the cosmic absorption.

\begin{figure}[htbp]
\includegraphics[width=8cm]{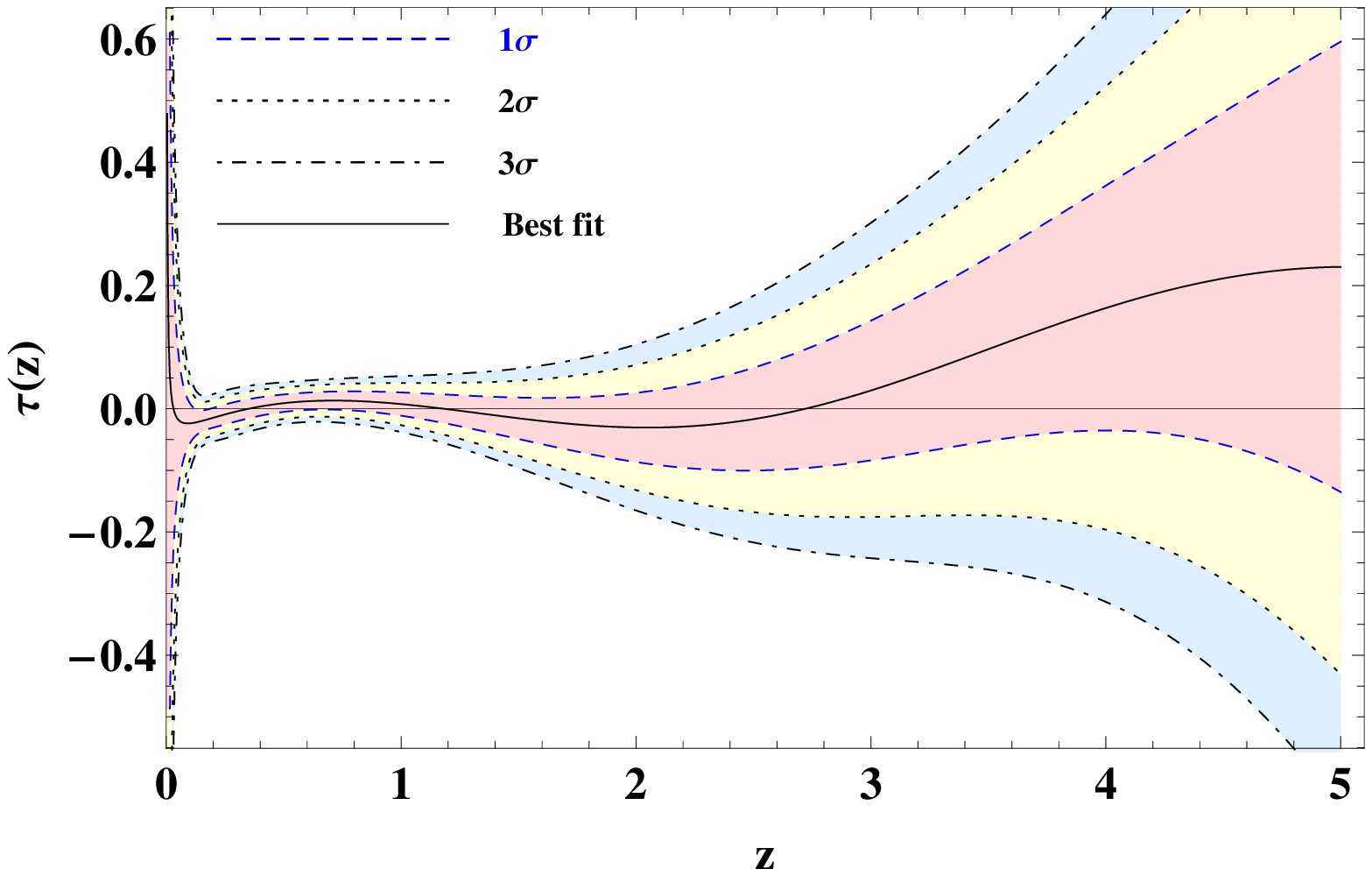}
\includegraphics[width=8cm]{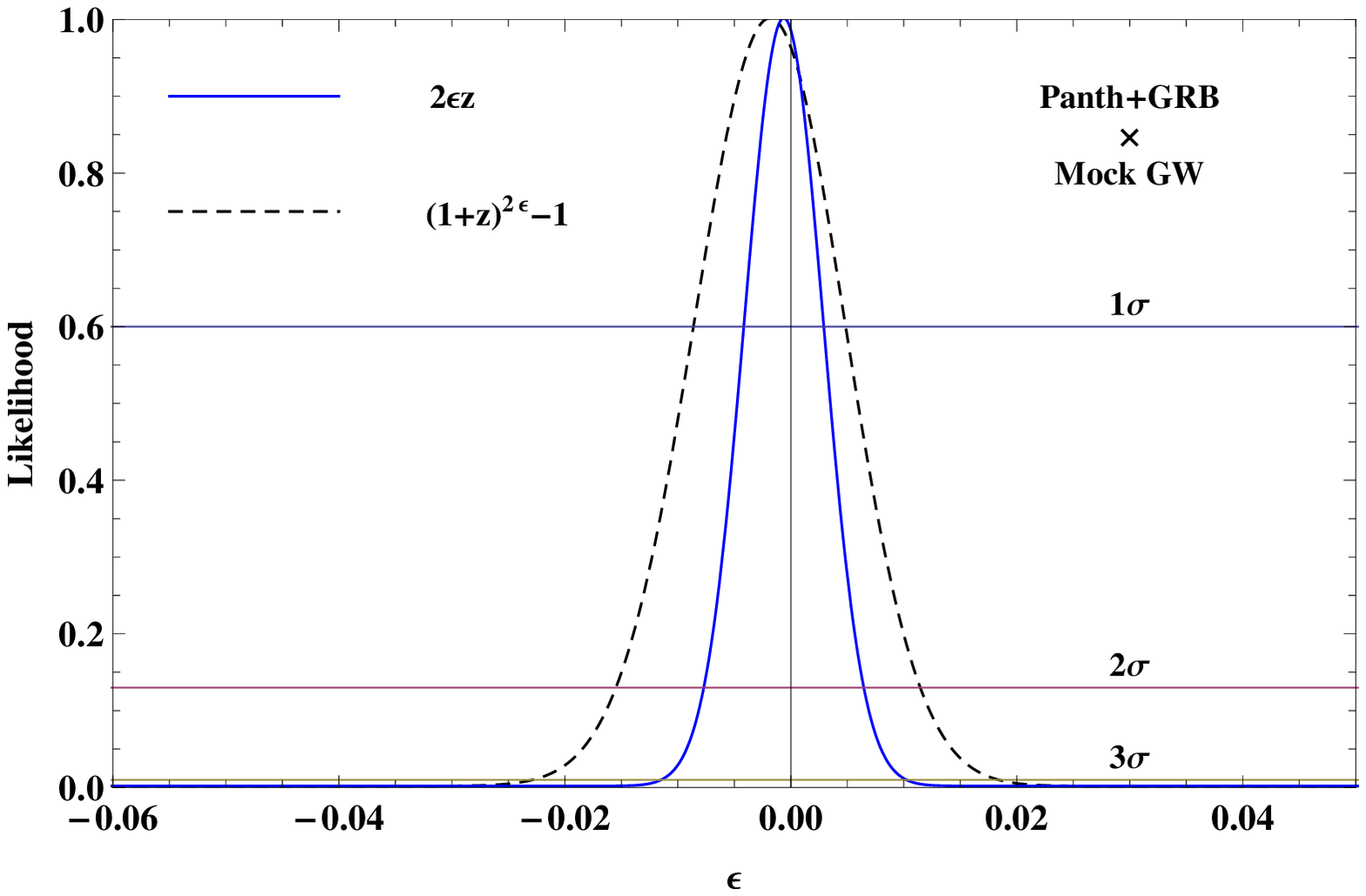}
\caption{\label{Figlikec} The cosmic optical depth function $\tau(z)$ (left panel) and the likelihood  distribution functions (right panel) obtained from the combination of `Pantheon-like' and `GRB-like' data with  $\Lambda$CDM.}
\end{figure}

Then the function of cosmic optical depth at any redshifts can be obtained through the following equation
\begin{equation}\label{opcity2}
{\tau(z)}= 2\ln {D_{\rm L,GW}(z) \over{D_{\rm L,\,EM}(z)}}
\end{equation}
 by comparing the continuous function of  LD $D_{\rm L}(z)$ reconstructed from the mock GW data with that from  the EM observational data. Here and latter, the subscript ``EM" represents the observed variable obtained from the electromagnetic wave measurements, such as the SNIa and GRB observations. Deviations from a transparent universe at any redshifts will be encoded in the function $\tau{(z)}$.
 The uncertainties of the optical depth from the observations, $\sigma_{\tau_{\rm obs}}$, can be obtained from following equation,
\begin{equation}
\label{SGL}
\sigma_{\tau_{\rm obs}}=2\sqrt{\left[\left({\sigma_{D_{\rm L,\,EM}(z)}\over{D_{\rm L,\,EM}(z)}}\right)^2+\left(\sigma_{D_{\rm L, GW}(z)}
\over{D_{\rm L, GW}(z)}\right)^2\right]}\, .
\end{equation}
The results are shown in the left panel of Fig.~\ref{Figlikec}. It should be noted that the constraint on cosmic opacity  is independent of any parametrization.

In addtion, we employ
two typical  parameterizations for the optical depth $\tau{(z)}$, i.e. $\tau(z)=2\epsilon z$ (P1)  and $\tau(z)=(1+z)^{2\epsilon}-1$ (P2), to detect cosmic opacity in the redshift region $0<z<5$.  While $\epsilon=0$, the Universe is transparent. Here, the observational data of opacity-free LD is from the simulated GW observation. To match the GW measurement with the EM observation at the same redshift and employ all GW data to test the cosmic opacity,  the opacity-dependent one is from the function of LD reconstructed from EM measurements with Gaussian process. The probability density of $\epsilon$ can be obtained  with the expression $P(\epsilon)=A\,{\rm exp}[-\chi^2(\epsilon)/2]$, where, $A$ is a normalized coefficient and $\chi^2(\epsilon)$ has the form
\begin{equation}
\label{chi3}
\chi^{2}(\epsilon) = \sum\frac{{\left[\tau(z,\epsilon)-
\tau_{\rm obs}\right] }^{2}}{\sigma^2_{\tau_{\rm obs}}}\,.
\end{equation}
The results are shown in the right panel of Fig.~\ref{Figlikec}.

\begin{figure}[htbp]
\includegraphics[width=8cm]{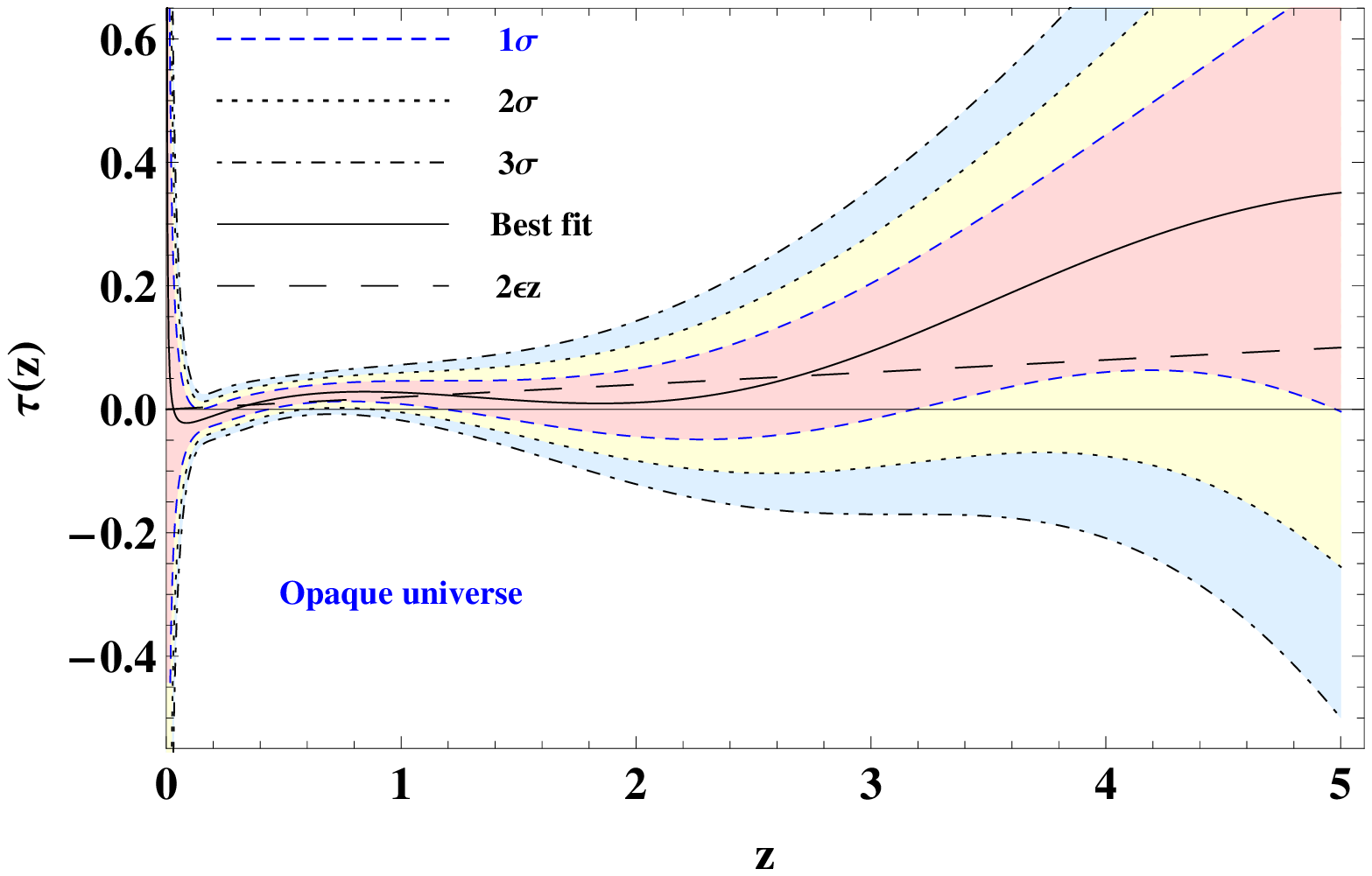}
\includegraphics[width=8cm]{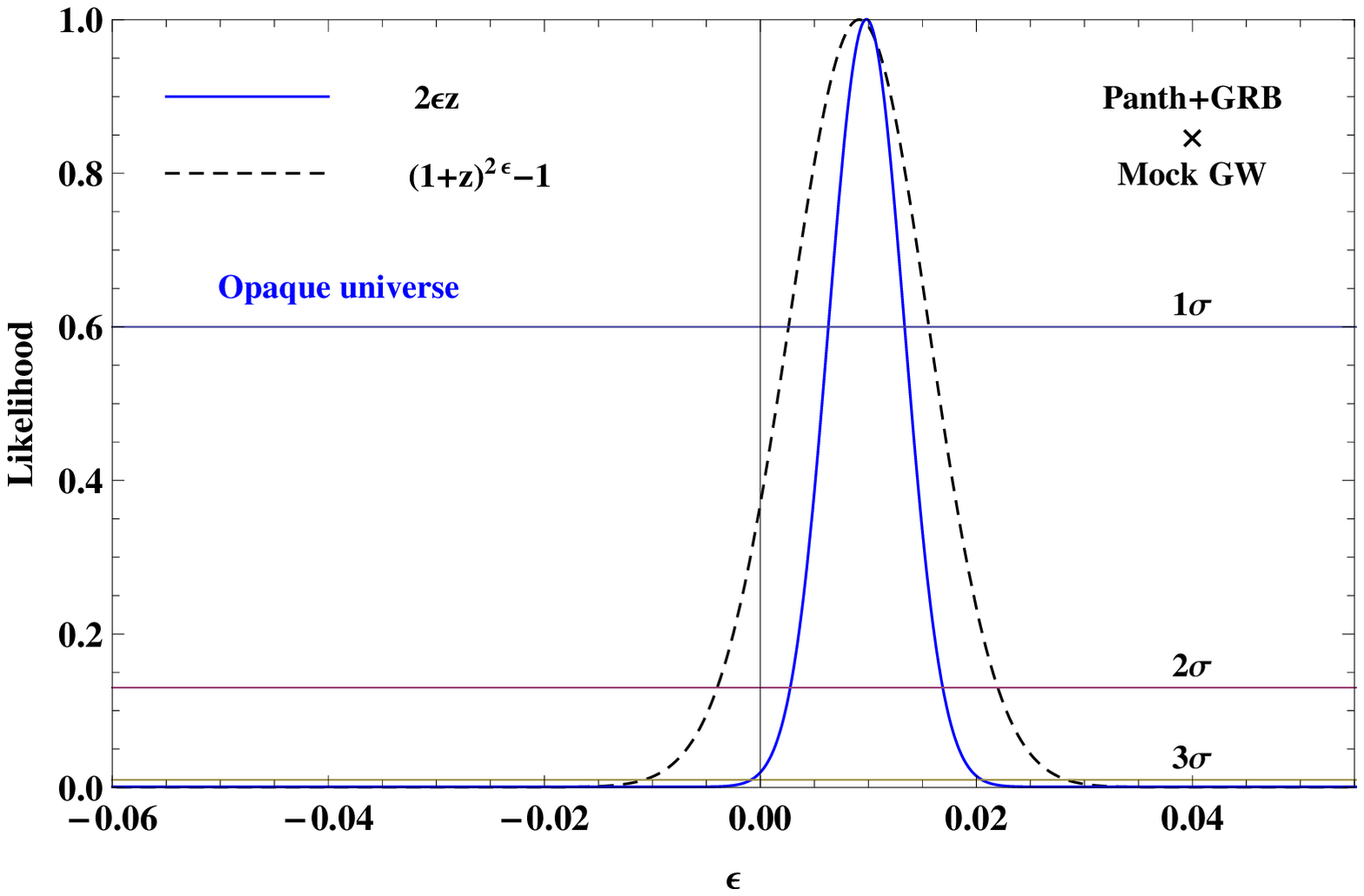}
\includegraphics[width=8cm]{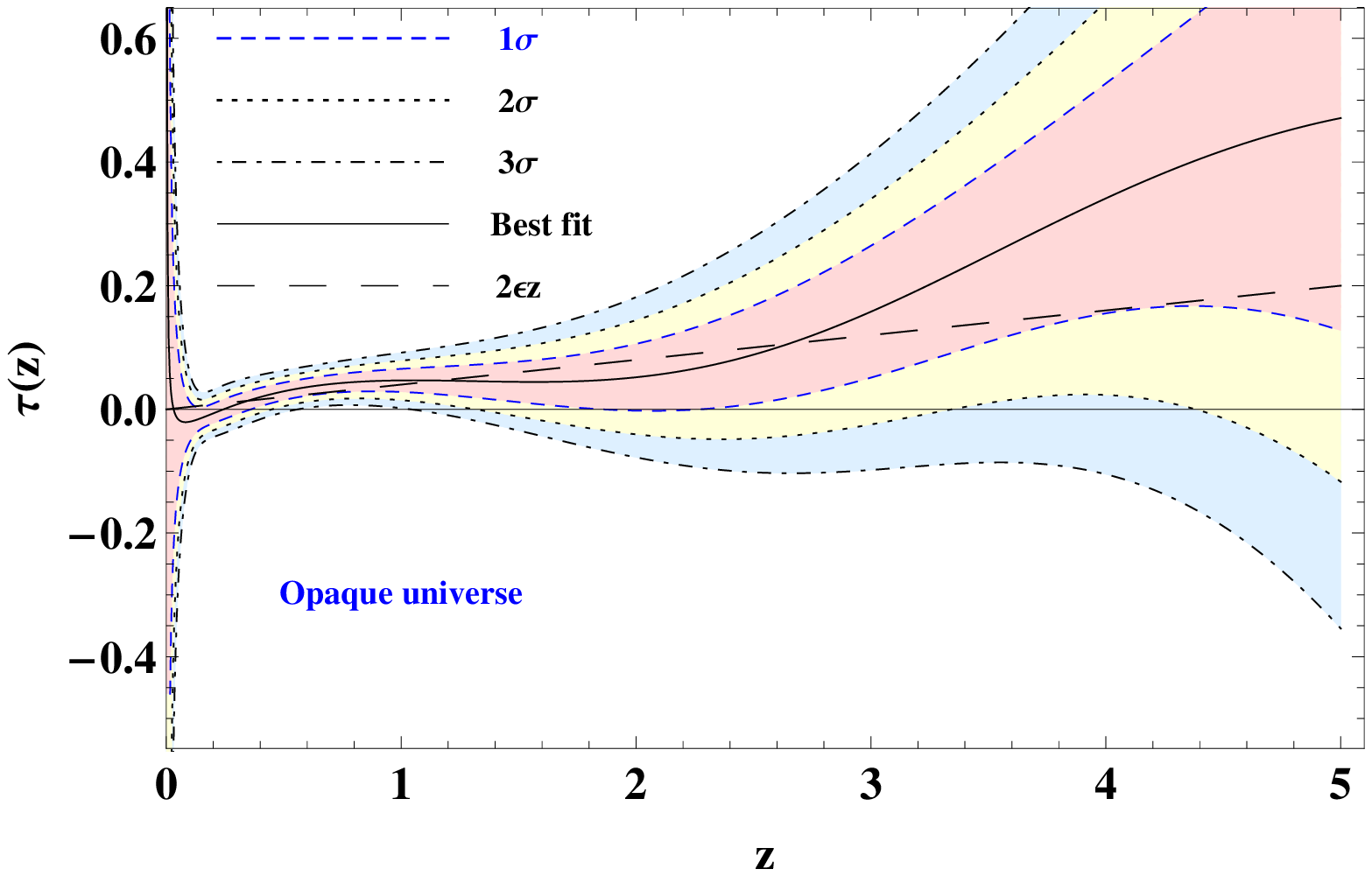}
\includegraphics[width=8cm]{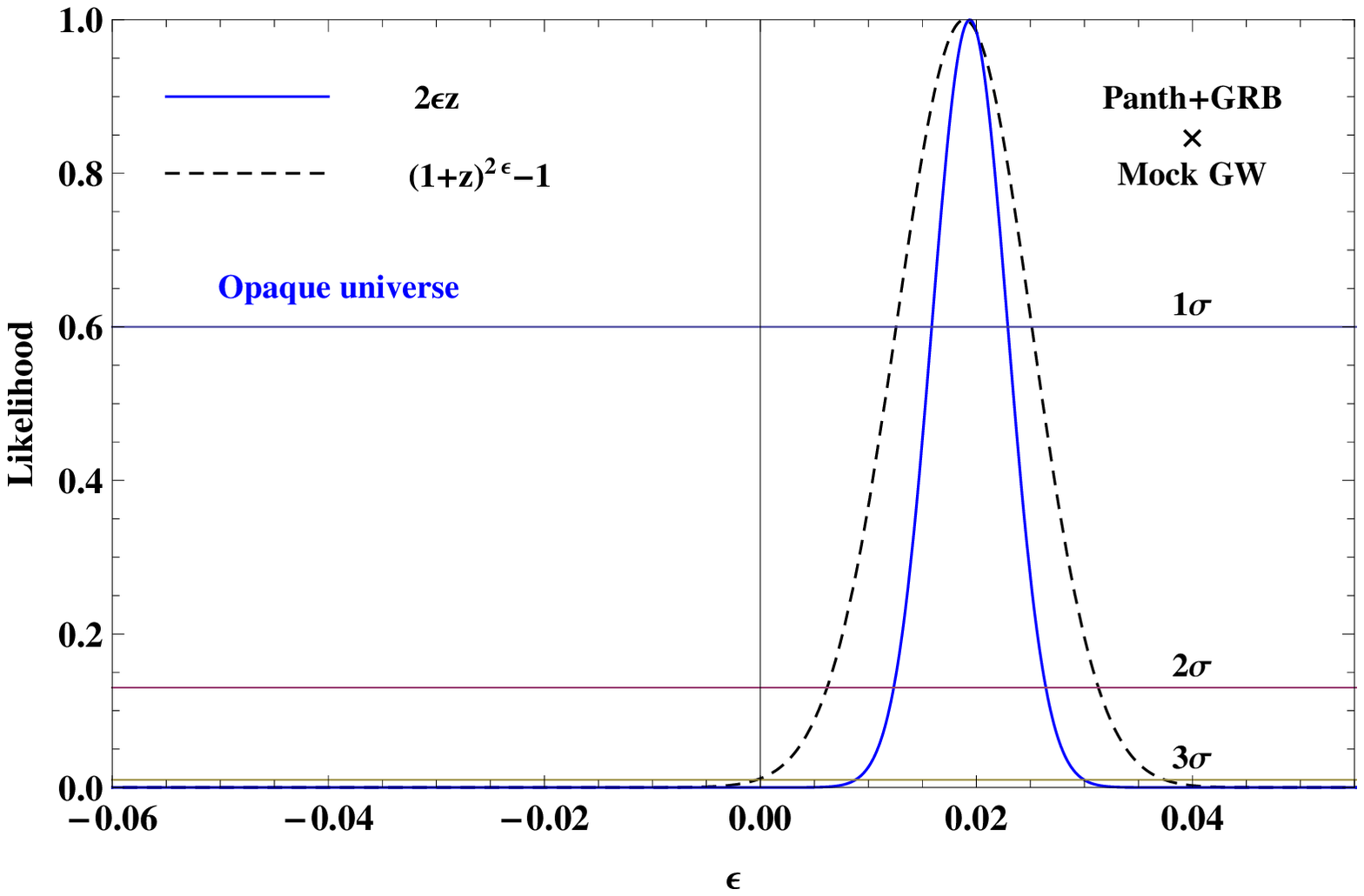}
\caption{\label{figtao} The cosmic optical depth function $\tau(z)$ (left panel) and the likelihood  distribution functions (right panel) obtained from the combination of `Pantheon-like' and `GRB-like' data with a certain model of opacity in parametrization $\tau(z)=2\epsilon z$ with $\epsilon=0.01$ (top panel) and $\epsilon=0.02$ (bottom panel), respectively.}
\end{figure}

Furthermore, in order to verify the effectiveness of this method, we also simulated SNIa data and GRB data by assuming that the universe is opaque in the case of optical depth function $\tau(z)=2\epsilon z$. The values of $\epsilon$ are set to $\epsilon=0.01$ and $0.02$, respectively. Then, the cosmic opacity are tested, and the results from the tests are shown in Fig.~\ref{figtao}.
\section{Results and analyses}
For the test on the spatial homogeneity of cosmic transparency in Fig.~\ref{Figlikec},  it can be seen that the best-fit value of cosmic optical depth $\tau(z)$ evolves  with the increase  of the redshift. No deviation from a  transparent universe is found  at $1\sigma$ CL. Our results are similar to those obtained in Refs.~\cite{Chen2012,fuxiangyun20192} in the range of SNIa measurement redshifts, in which the best fit value of cosmic opacity varies with  the increase of the redshift. In addition, the value of cosmic optical depth $\tau(z)$  is constrained between  $-0.09$ and $0.05$ at $1\sigma$ CL  in the redshift region $0.1<z<2.0$, and it is constrained between $-0.14$ and $0.60$ in the redshift region $2.0<z<5.0$.  It suggests that the potentiality of observational data is relatively weak to probe the spatial homogeneity of cosmic opacity at the high redshift. As presented in Section~\ref{method}, due to a small number and large uncertainty of GRB observations in this redshift region,  the uncertainty of LDs reconstructed from the GRB data  is greater than that of LDs reconstructed from the GW measurements.  Therefore, the improvements of  electromagnetic observations in the high redshift region, including the improvements of data quantity and quality,  will impose more stringent constraints on the test of cosmic opacity. The figure also shows that  the cosmic optical depth $\tau(z)$ is divergent in the redshift range $z<0.1$, and this bias might result from the absence of the mock GW data in this redshift region.

For the case of tests on cosmic opacity with parameterizations, as shown in Tab.~\ref{likelihood1}, we also present a  comparison between our forecast results and the previous ones.
Compared  the error bars of the constraint on cosmic opacity with the previous ones from the SNIa data in the lower redshift range $0<z<2$,  our results are  at least 50\% less than those obtained from the simulated ET GW data and the JLA SNIa data  in Refs.~\cite{JUN-JIE WEI2019,Jing-Zhao Qi2019,fuxiangyun20192}, and they are at least 30\% less than those from the simulated ET GW data and Pantheon SNIa compilation in Ref.~\cite{fuxiangyun20192}.
 Therefore, the future GW measurements can be considered as an important tool to probe cosmic opacity. For the results obtained from parameterizations for the optical depth $\tau(z)$, it is easy to see that parametrization P1 may offer more stringent constraint on cosmic opacity than P2.
 \begin{table}[htp]
\begin{tabular}{c c c c }
\hline\hline\\
\scriptsize{Data }  & \   $\epsilon\,\,\,( P_1)$\ \ &$\epsilon \,\,\,( P_2)$ \\
\hline
\scriptsize{ET $\times$ 1048 Pantheon + 137 GRB$^\ast$ } &  \scriptsize{${-0.001{\pm{0.004}\pm{0.007}}}$} & \scriptsize{${-0.002{\pm{0.007}\pm{0.013}}}$} \\
\scriptsize{ET $\times$ 1048 Pantheon + 137 GRB$^\ast\lozenge$ } &  \scriptsize{${0.010{\pm{0.004}\pm{0.007}}}$} & \scriptsize{${0.009{\pm{0.006}\pm{0.012}}}$} \\
\scriptsize{ET $\times$ 1048 Pantheon + 137 GRB$^\ast\lozenge$ } &  \scriptsize{${0.019{\pm{0.004}\pm{0.007}}}$} & \scriptsize{${0.018{\pm{0.006}\pm{0.012}}}$} \\
\scriptsize{ET $\times$ 740 JLA  \tiny{(A)}~\cite{fuxiangyun20192}} &  \scriptsize{${0.004{\pm{0.008}}}$} & \scriptsize{${0.002{\pm{0.010}}}$} \\
\scriptsize{ET $\times$ 740 JLA \tiny{(B)}~\cite{fuxiangyun20192}} & \scriptsize{${0.009{\pm{0.011}}}$}&\scriptsize{ ${0.008{\pm^{0.014}_{0.012}}}$}   \\
\scriptsize{ET $\times$ 1048 Pantheon \tiny{(B)}~\cite{fuxiangyun20192}} & \scriptsize{${0.005{\pm{0.006}}}$}&\scriptsize{ ${0.006{\pm{0.008}}}$}   \\
\scriptsize{ET $\times$ 1048 Panthoen \tiny{(B)} \cite{JUN-JIE WEI2019}} & \scriptsize{${0.004\pm 0.026}$}& $\Box$   \\
\scriptsize{ET $\times$ 740 JLA \tiny{(B)} \cite{Jing-Zhao Qi2019}} & \scriptsize{${0.002\pm 0.035}$}&\scriptsize{ ${-0.006\pm 0.053}$}   \\
\scriptsize{ET $\times$ 1048 Panthoen \tiny{(B)} \cite{Jing-Zhao Qi2019}} & \scriptsize{${0.009\pm 0.016}$}&\scriptsize{ ${0.015\pm 0.025}$}   \\
\scriptsize{581 SNIa + 19 H(z) $ \times$ $\Lambda$CDM \tiny{(A)} \cite{Holanda2014}} & \scriptsize{${0.02\pm0.055}$}~$\triangle$& $\Box$\\
\scriptsize{59 GRB + 19 H(z) $ \times$~$\Lambda$CDM \tiny{(A)} \cite{Holanda2014}} & \scriptsize{${0.06\pm^{0.18}_{0.18}}$}~$\triangle$& $\Box$\\
\hline\hline
\end{tabular}
\caption{Constraints on $\epsilon$ represented by the best fit value at $1\sigma$ or $2\sigma$ CL for each data set.  The superscripts `$\ast$' represent the results  obtained from the Gaussian process in this work, the symbol `$\lozenge$' refers to the results obtained from a certain model of opacity, A and B represent the cosmological model-dependent method and the model-independent method, respectively, and the triangle symbols `$\triangle$' denote the parametrization is adopted in the form $\tau(z)=\epsilon z$.}
\label{likelihood1}
\end{table}

For the test on the spatial homogeneity of the optical depth $\tau(z)$ from a certain model of cosmic opacity while $\epsilon=0.01$,  an opaque universe can be found at $2\sigma$ CL in the redshift region $0.46<z<0.82$, as shown in the left-top panel of Fig.~\ref{figtao}. It is worth mentioning that the test on cosmic opacity obtained directly from the observational data is independent of any parametrization. For the test from the parametrization of the optical depth $\tau(z)$, an opaque universe can be marginally found at $3\sigma$ CL for the parametrization of P1, as shown in the right-top panel of this figure. While $\epsilon=0.02$, it can be seen from bottom panel of Fig.~\ref{figtao} that an opaque universe can be found at $3\sigma$ CL. It suggests that future gravitational wave data can be used as an effective tool to identify an opaque universe.

\section{conclusion}
In the past few years, some astronomical observations, such as the luminosity distance (LD) of SNIa, baryon acoustic oscillations, gas mass fraction, angular diameter distances from galaxy clusters, and Hubble parameter measurements have been used to probe cosmic opacity. As a gravitational wave propagates freely in a perfect fluid without any absorption and dissipation, the measurement of the luminosity distance  to a GW source provides us with the opacity-free distance to probe cosmic opacity. More recently, confronting the mock GW measurements to the SNIa compilations, some tests on cosmic opacity have been performed to explore the potentiality of  future GW measurements.  However, due to the limitation from the redshift distributions of observational data, most of the tests are limited in the redshift region $0<z<2$, and the cosmic opacity in the redshift region beyond the SNIa observations remains unexplored.

The third-generation  GW detector, i.e., the Einstein Telescope (ET), is expected to detect GW signals up to redshift range $z\sim 5$. Thus, future GW measurements will provide us with an opportunity to probe cosmic opacity at high redshifts.  In this work,  using  the Gaussian process, we explore the potentiality of  future gravitational wave
measurements to probe the cosmic opacity at high redshift through comparing  LD from GW with that from the combination of the SNIa  and GRBs compilations.  One thousand GW data points are simulated from the future ET.
To dodge any impacts of systematics while combining the SNIa and GRB data sets,  the `SNIa-like' and `GRBs-like' data are simulated from the actual Pantheon SNIa compilation and the latest GRB catalogue of Ref.~\cite{Amati2019} in the $\Lambda$CDM cosmological model. 137 GRB data points in the redshift region $1<z<5$ are added to the Pantheon compilation. The purpose of using the Gaussian process is to reconstruct the continuous function of LD from the future GW data or the combination of SNIa and GRB data sets. A non-parametric method are proposed to probe  the spatial
homogeneity of the cosmic transparency at any redshifts  through comparing the LD reconstructed from GW measurements  with that reconstructed from SNIa and GRB data.
Furthermore, using the simulated GW data and the continuous function of LDs reconstructed  from the combination of SNIa data and GRB data, the cosmic opacity is tested with two types of parameterizations for the optical depth.  All of the available GW data points  can be used to test cosmic opacity. To test the effectiveness of the method in this work, SNIa data and GRB data are also simulated  with a certain cosmic opacity model  to probe the cosmic spatial homogeneity and the cosmic opacity.

For the test on the spatial homogeneity of cosmic opacity, the results show that the best fit value of cosmic opacity varies as the redshift increases, and no deviation from a  transparent universe is found  at $1\sigma$ CL.  Due to the large uncertainty of the GRB observation in the high redshift region and the small amount of data, the potentiality of astronomical observation to test cosmic opacity in high redshift region is relatively weaker than that in the low redshift region dominated by the SNIa observations. It suggests that improvements from the electromagnetic observations at high reshifts, including the improvements of data quantity and quality, would lead to more stringent constraints on the cosmic opacity. Results, obtained from the test with parameterizations for the optical depth, show that the error bar of constraint on cosmic opacity can be reduced to $\sigma_{\epsilon}\sim 0.004$ at $1\sigma$ CL. The uncertainty of the constraint is  less than the previous results obtained from the simulated GW data, JLA SNIa data, and Pantheon SNIa data.  The tests from a certain model of opacity show that the GW measurements in the future may be used as an effective tool  to probe cosmic opacity.
Therefore, it can be concluded that, given the GW detector can be carried out as expected by the programme of the ET, future measurements of GW may not only provide us with an opportunity to investigate  the spatial homogeneity of the cosmic transparency at high redshifts,  but also play an important role in probing cosmic opacity.

\begin{acknowledgments}
We very much appreciate helpful comments and suggestions from anonymous referees.  We are also thankful for helpful discussion from Puxun Wu and  Zhengxiang Li.
This work was supported by the National Natural Science Foundation of China
under Grants No. 11147011, No. 11405052, No. 11465011, and No. 11865011;  the
Hunan Provincial Natural Science Foundation of China under Grant No. 12JJA001;
the Foundation of Department of science and technology of Guizhou
Province of China under Grant No. J [2014] 2150; and the Foundation of
the Guizhou Provincial Education Department of China under Grant
No. KY [2016]104.

\end{acknowledgments}


\end{document}